# Timing and Code Size Optimization on Achieving Full Parallelism in Uniform Nested Loops

Y. Elloumi, M.Akil and M.H. Bedoui

**Abstract**- Multidimensional Retiming is one of the most important optimization techniques to improve timing parameters of nested loops. It consists in exploring the iterative and recursive structures of loops to redistribute computation nodes on cycle periods, and thus to achieve full parallelism. However, this technique introduces a large overhead in a loop generation due to the loop transformation. The provided solutions are generally characterized by an important cycle number and a gr eat code size. It represents the most limiting factors while implementing them in embedded systems.

In this paper, we present a new Multidimensional Retiming technique, called "Optimal Multidimensional Retiming" (OMDR). It reveals the timing and data dependency characteristics of nodes, to minimize the overhead. The experimental results show that the average improvement on t he execution time of the nested loops by our technique is 19.31% compared to the experiments provided by an existent Multidimensional Retiming Technique. The average code size is reduced by 43.53% compared to previous experiments.

——————————— ◆ ———————————

## 1 INTRODUCTION

The design of real time systems should respect many constraints such as the execution time and code size, which require using optimization techniques. The Retiming presents one of these techniques which can be used to add and remove registers in order to provide a more efficient circuit [1].

The increased complexity of such application leads to the frequent use of a nested iterative and recursive loops. Such applications can be modeled as "Multidimensional Data Flow Graph" (MDFG). The standard software pipelining techniques can only be used to optimize a one-dimensional loop. When they are applied to optimize nested loops, the performance improvement is very limited [8].

Other works are proposed to offer an optimization technique taking advantage of the multiple nested loops, which is called "Multi-Dimensional Retiming" (MDR). It aims achieving full parallelism of uniform nested loops. It consists in scheduling the MDFG with the minimum cycle period and modifying the execution order of nodes, such as each one is executed in a separate cycle.

But, Achieving full parallelism requires adding a large code overhead [2],[3]. It dramatically increases the whole code size of the provided MDFG. Furthermore, this extra code requires a significant cycle number to be executed outside the loop body. Thus, the provided solution does not allow achieving an application with an adequate execution time and a code size. It represents a limiting factor to implement the provided MDFG in a real-time embedded system.

We propose in this paper a new technique of MDR, called "Optimal Multidimensional Retiming". It allows redistributing optimally the nodes on cycle periods, while scheduling the MDFG with the minimal cycle period. This technique significantly allows optimizing the number of period cycles (notably the execution time) and the code size, by exploring the execution time and data dependency between nodes belonging to the MDFG. Thus, it provides enhanced solutions, compared to the existent techniques.

The rest of the paper is organized as follow. In section 2, we give an overview of MDFG formalism. In section 3, we list the existent MDR techniques and their constraints and limits. In section 4, we present the theory of the "Optimal Multidimensional Retiming" technique by describing the principles and basics concepts, and proposing the correspondent algorithms. Experimental results are presented in section 5, followed by concluding remarks in section 6.

## 2 MULTIDIMENSIONAL DATA FLOW GRAPH

The Multidimensional Data Flow Graph (MDFG) is an extension of the classic data flow graph that allows to represent a nested iterative and recursive structures. It is modeled by a node-weighted and edge-weighted directed graph such as $G = (V, E, d, t)$, where V is the set of computation nodes, $E \subseteq V \times V$ is the set of edges, and $d(e_i)$ is a function from E to $Z^n$, representing the multidimensional delay between two nodes, where n is the number of dimensions (loops), and $t(v_j)$ is a function from V to the positive integers, representing the computation time of the node $v_j$.

For a MDFG with n dimensions, each edge $e : v_i \rightarrow v_j$ is characterized by a delay where $d(e) = (c_1, c_2, \ldots, c_n)$. The value $c_k$ represents the difference between the execution iteration of $v_j$ and the execution iteration of $v_i$ of the loop $k$.

We show in Fig.1.a a two-dimensional Data Flow Graph (2DFG) corresponding the Wave Digital Filter described in Algorithm 1, which is composed of two nested loops. The execution of each node in V exactly represents one iteration, which is the execution of one instance of the loop body. Each edge belonging to the 2DFG shown in Fig.1.a is labeled by a delay $d(e) = (d.x, d.y)$. Both terms « $d.x$ » and « $d.y$ »

————————————————

- *Y. Elloumi is a P.H.D. student in Paris-Est University, 93162 Noisy le Grand Cedex, France.*
- *M. Akil is Professor at computer science department, ESIEE, Paris, 93162 Noisy le Grand Cedex, France.*
- *M.H. Bedoui is Professor in faculty of medicine of Monastir, University of Monastir, 5019, Monastir, Tunisia,*

represent the difference between the iteration number executing $v_j$ and the iteration number executing $v_i$, in the outermost loop as well as in the innermost loop [9].

For an edge $e : v_i \rightarrow v_j$, the delay $d(e) = (0, x)$ consists in the execution of $v_i$ and $v_j$ in the same iteration of the outermost loop. For the innermost loop, if the node $v_i$ is executed in the iteration $k$, the node $v_j$ is executed in the

ALGORITHM 1
WAVE DIGITAL FILTER

```
0:  For i from 0 to m do
1:    For j from 0 to n do
2:      D(i,j)= B(i-1 , j+1) × C(i-1 , j-1)
3:      A(i,j)= A(i,j) × 5
4:      B(i,j)= A(i,j) + 1
5:      C(i,j)= A(i,j) + 2
6:    End for
7:  End for
```

iteration$(k - x)$. An edge with zero delay $d(e) = (0,0)$ represents a data dependency in the same iteration, such as the edges $D \rightarrow A$, $A \rightarrow B$ and $A \rightarrow C$ as shown in Fig. 1.a.

We use the notation $v_i \xrightarrow{e} v_j$ to indicate that $e$ is an edge from $v_i$ node to $v_j$ node, and $v_i \xRightarrow{p} v_j$ to mean that $p$ is a path from $v_i$ node to $v_j$ node. The delay vector of a path $p: v_i \xrightarrow{e_m} v_{i+1} \xrightarrow{e_{m+1}} ... \xrightarrow{e_n} v_j$ is $d(p) = \sum_{k=m}^{k=n} d(e_k)$ and the total computation time of a path $p$ is $t(p) = \sum_{k=i}^{k=j} t(v_k)$. The period during which all computation nodes in iteration are executed, according to existing data dependencies and without resource constraints, is called a cycle period. The cycle period $C(G)$ of an MDFG is the maximum computation time among paths that have a zero delay. For example, assuming that each node is executed in one time unit $t(A) = t(B) = t(C) = t(D) = 1$, the MDFG of Fig.1.a has $C(G) = 3$. It can be measured through the paths $p: D \rightarrow A \rightarrow B$ or $p: D \rightarrow A \rightarrow C$, as shown in the iteration scheduling illustrated in Fig.1.b. Each set of nodes belonging to the same iteration are modeled by a different motif.

The execution pattern of a nested loop can be illustrated by iteration space as shown in Fig.2.a. Each cell in the iteration space is a copy of the MDFG. The marked cell, labeled by (0,0), is the first iteration to be executed. This graph is transformed on an acyclic graph, called cell dependency graph (CDG), allowing to show clearly the execution sequence of a nested loop. The CDG of an MDFG G illustrates the dependencies between copies of nodes representing the MDFG G, such as the CDG shown in Fig.2.b which corresponds to the MDFG G in Fig.1.a. A node in CDG is a computational cell that represents a complete iteration. The CDG of a nested loop is bounded by the loop indexes.

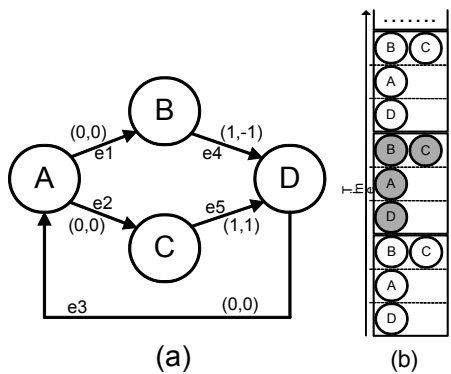

(a)      (b)

Fig.1. (a) MDFG of Wave Digital Filter; (b) Iteration scheduling of the MDFG in Fig.1.a.

A schedule vector s defines a sequence of execution in the cell dependency graph. The CDG shown in Fig.2.b, can be executed by a row-wise execution sequence, i.e., the schedule vector $s = (1,0)$. A legal MDFG $G = (V, E, d, t)$ is realizable if there exists a schedule vector s for the cell dependency graph in respect to G; i.e., $s \times d(e) \geq 0, e \in E$, and no cycle exists in its corresponding CDG. Note that delay vectors $(0,1)$ and $(0,-1)$ are both legal in respect to the schedule vector $s = (1,0)$, but they create cycles in cell dependency graph.

## 3 MULTIDIMENSIONAL RETIMING

### 3.1 Principles

The retiming technique consists in redistributing delays in the graph. This technique can be applied on a data flow graph to minimize the cycle period in a polynomial time. The delays are moved around in the graph in the following way: a delay unit is drawn from each of the incoming edges of $v$, and then added to each of the outgoing edges of $v$, or vice versa [1]. In the case of MDFG, it consists in redistributing the execution of nodes on the iterations. The retiming vector $r(u)$ of a node $u \in G$ represents the offset between the original iteration containing $u$, and the one after retiming. Note that the retiming technique preserves data dependencies of the original MDFG. Therefore, we have $d_r(e) = d(e) + r(u) - r(v)$ for every edge and $d_r(l) = d(l)$ for every cycle $l \in G$. After retiming, the execution of the node u in the iteration i is moved to the iteration $i - r(u)$.

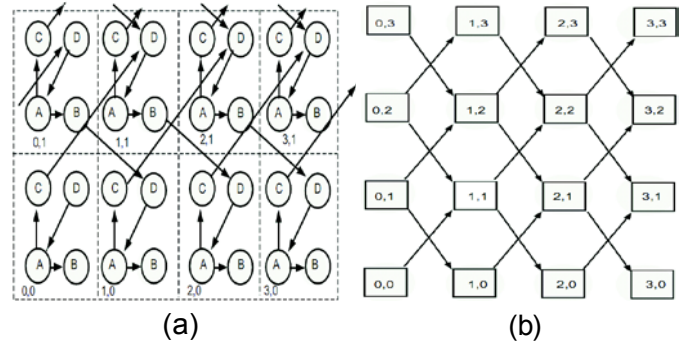

(a)      (b)

Fig.2. (a) Iteration space of the MDFG in Fig.1; (b) The cell dependency graph.

We show in Fig.3.a the MDFG $G_r = (V, E, d_r, t)$ of the wave digital filter after applying the retiming function $r(D) = (0,1)$. When a delay is pushed through node $D$ to its outgoing edge as shown in Fig.3.a, the actual effect on the Algorithm 2 of the new MDFG is that the $i^{th}$ copy of $D$ is shifted up and is executed with $(i - (0,1))^{th}$ copy of nodes $A$, $B$, and $C$. The original zero-delay edge $D \rightarrow A$ in Fig.1.a now has a delay $(0,1)$ after retiming as shown in Fig.3.a. Node $D$ in the new loop body has not any data dependency with other nodes executed in the same cycle. So, node $D$ can be executed in parallel to node $A$, as shown in the iteration scheduling of Fig.3.b. Thus, the cycle period is reduced from three to two time units.

In fact, every retiming operation corresponds to a software pipelining operation. When $r(u)$ delay units are pushed through a node u, every copy of this node is moved by $r(u)$ iterations. Hence, a new iteration consists in redistributing the execution nodes into different iterations.

ALGORITHM 2
WAVE DIGITAL FILTER AFTER RETIMING BY THE FUNCTION
R(D)=(0,1)

```
0:  For i from 0 to m do
1:    D(i,0) = B(i-1 , 1) × C(i-1 ,-1)
2:    For j from 0 to n-1 do
3:      D(i,j+1) = B(i-1 , j+2) × C(i-1 , j)
4:      A(i,j) = D(i,j) × 5
5:      B(i,j) = A(i,j) + 1
6:      C(i,j) = A(i,j) + 2
7:    End for
8:    A(i,n) = D(i,n) × 5
9:    B(i,n) = A(i,n) + 1
10:   C(i,n) = A(i,n) + 2
11:End for
```

Some nodes are shifted out of the loop body to provide the necessary data for the iterative process, which is called prologue. Correspondingly, some nodes will be executed after the loop body to complete the process, which is called epilogue.

Using MD retiming function r, we can trace the pipelined nodes and also measure the size of the prologue and epilogue. For node v with retiming $r(v) = (i, j)$, there are i copies of node v appearing in the prologue of outer loop, and j copies of node v in the prologue of the innermost loop. The

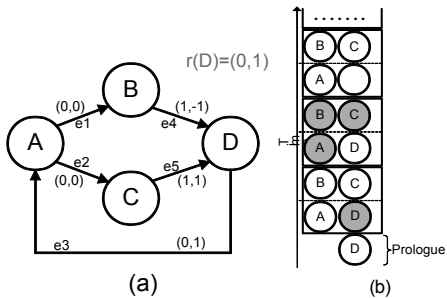

Fig.3. (a) MDFG of algorithm 2; (b) Iteration scheduling of the MDFG in Fig.3.a number of copies of a node in the epilogue can also be derived in a similar way. The iteration space of the retimed MDFG shown in Fig.4.a with retiming $r(D) = (0,1)$ clearly shows that one copy of node D is pushed out of the loop body on j-dimension, and becomes prologue for the innermost loop. The corresponding cell dependency graph is shown in Fig.4.b.

It is known that an MDFG can always be fully parallelized by applying successively the MD retiming functions $r(D) = (0,2)$ and $r(A) = (0,1)$, which is illustrated in Fig.5.a. We note that the retimed MDFG has non-zero-delay on each edge. It implies that the nodes belonging to the same iteration in the original loop body are distributed into three cycle periods. The MDFG is then scheduled with the minimal cycle period equal to one time unit, as schematized the iteration scheduling of Fig.5.b.

To achieve a realizable MDFG after retiming, the legality condition, $s \times d(e) \geq 0$, has to be satisfying, and there should not exist any cycle in the cell dependency graph of the MDFG.

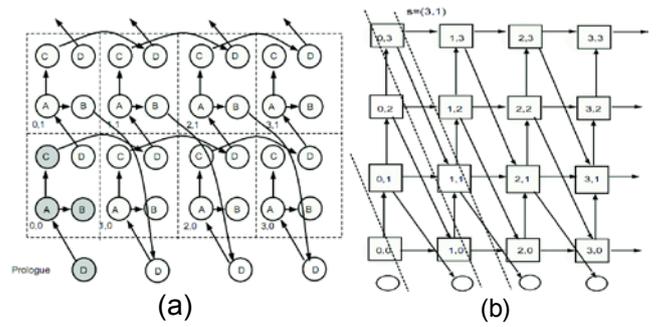

Fig.4. (a) The iteration space the retimed MDFG in Fig.3; (b) The cell dependency graph.

Hence, the MDR technique aims to transform a realizable MDFG G on MDFG $G_r$ in a way that $G_r$ is still realizable. Using such concepts, the basic conditions for legal multidimensional retiming are defined in the following

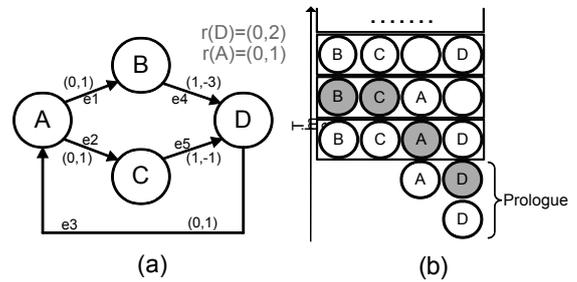

Fig.5. (a) Fully parallelized graph of the MDFG in Fig.1; (b) Iteration scheduling of the MDFG in Fig.5.a lemma [2].

**Lemma 1.** *Let* $G = (V, E, d, t)$ *be a realizable MDFG, r a multidimensional retiming, and s a schedule vector for the retimed graph* $G_r = (V, E, d_r, t)$, *then*

1. *for any path , we have* $d_r(p) = d(p) + r(u) - r(v)$
2. *for any cycle* $l \in G$ *we have* $d_r(l) = d(l)$
3. *for any edge* $u \xrightarrow{e} v$, $d_r(e) \times s \geq 0$
4. *there is no cycle in the DG equivalent to the MDFG G.*

The selection of a legal multidimensional Retiming function is based on the edge delay of the MDFG. The approach proposed in [2],[3] consists in defining a scheduling subspace S for a realizable MDFG $G = (V, E, d, t)$. It represents the space region where there exist schedule vectors that realize $G$; i.e., if schedule $s \in S$ then $s \times d(e) \geq 0$ for any $e \in E$. In fact, the multidimensional retiming technique means to decrease zero-delay edges. Thus, a strictly positive scheduling subspace $s^+$ is the set al all vectors $s \in S$ where $s \times d(e) > 0$ for every $d(e) \neq (0,0, ... ,0)$. The method of predicting a legal multidimensional retiming is introduced in the next theorem.

**Theorem 1.** *let* $G = (V, E, d, t)$ *be a realizable MDFG,* $s^+$ *a strictly positive scheduling sub-space of G, s a scheduling vector in* $s^+$, *and* $u \in V$ *a node with all incoming edge having nonzero delay. A legal MD retiming r of u is any vector orthogonal to s.*

### 3.2 Multidimensional Retiming techniques

We describe in this section the existent multidimensional retiming techniques. They are characterized by achieving full parallelism by providing the MDFG with no zero-delay edge [2],[3],[7].

  a) Incremental Multidimensional Retiming

This technique is based on selecting a set of nodes that can be retimed by the same multidimensional retiming function, as described in the following corollary.

**Corollary 1.** *Given a realizable MDFG $G = (V, E, d, t)$, $s^+$ a strictly positive scheduling sub-space of G, s a scheduling vector*

ALGORITHM 5
CHAINED MULTIDIMENSIONAL RETIMING

| |
|---|
| Input : a realizable MDFG G =(V,E,d,t) |
| Output : a realizable MDFG $G_r$=(V,E,$d_r$,t) without d(e)=(0,0, … ,0) |
| 0: Begin |
| 1: Find a legal MDR function r as described in steps 2 and 3 in algorithm 3 |
| 2: Provide the multi-chain graph and the maximal length of chain K, as indicated in algorithm 4 |
| 3: For each node v with label i do |
| 4:     Apply the MDR function (k-i)× (r) |
| 5: End for |
| 6: End |

in $s^+$, and an MD retiming function r orthogonal to s, if a set $X \subseteq V$ has all incoming edges nonzero, then $r(X)$ is a legal MD retiming.

Thus, this technique consists in defining a schedule vector s as described in definition 1, and chooses an MDR function orthogonal to s. This chosen function is applied to each node respecting the previous corollary. Those steps are repeated incrementally, until all zero-delay edges are transformed, as described in algorithm 3.

ALGORITHM 3
INCREMENTAL MULTIDIMENSIONAL RETIMING

| |
|---|
| Input : a realizable MDFG G =(V,E,d,t) |
| Output : a realizable MDFG $G_r$=(V,E,$d_r$,t) without d(e) = (0,0, … ,0) |
| 0: Begin |
| 1: While exist zero-delay edge in the graph Do |
| 2:     Find a scheduling vector s=(s.x,s.y) that s.x+s.y is minimum |
| 3:     Choose a MDR function |
| 4:     Apply the selected MDR function to any nodes that has all incoming edges with nonzero delays and at least one outgoing edge with zero delay |
| 5: End while |
| 6: End |

We apply the algorithm above to the 2DFG of Infinite

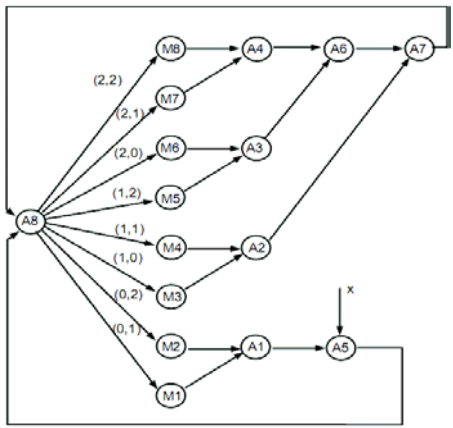

Fig.6. MDFG of IIR Filter.

Impulse Response filter (IIR) that is illustrated in Fig.6. It is composed by multiplier nodes assigned by $M_i$ and adder nodes assigned by $A_j$. Fig.7 illustrates the full parallelized MDFG of the IIR filter after applying the Incremental Multidimensional Retiming.

The steps of algorithm 3 are repeated four times, where in each one a different MD retiming function is applied. The fully parallelized MDFG is showed in Fig.7 where all edges are non-zero delay.

a) Chained Multidimensional Retiming [2],[3]

This technique allows obtaining the full parallelism solution by defining just one MDR function. It is based on the following corollary.

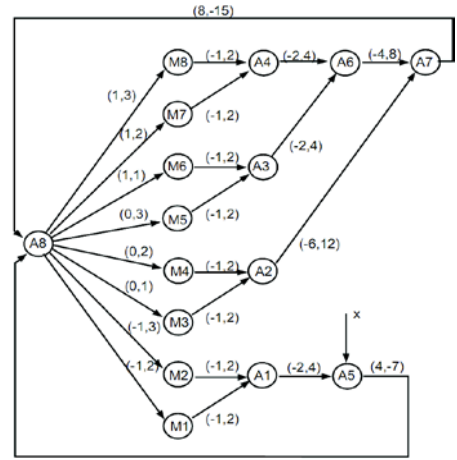

Fig.7. IIR Filter MDFG after Incremental MDR.

**Corollary 2.** *[2] Given $G = (V, E, d, t), S^+$ a strict positive scheduling subspace for G, s a scheduling vector $S^+$, a MD retiming function orthogonal to s, a set $X \subseteq V$ which all incoming edges nonzero, and an integer value $k > 1$, then $(k \times r)(X)$ is a legal MD retiming.*

Thus, it applies the MD retiming to successive nodes in a path where each node has a retiming function multiple of the selected retiming value smaller than its predecessor nodes.

ALGORITHM 4
MULTI-CHAIN GRAPH CONSTRUCTION

| |
|---|
| Input : a realizable MDFG G =(V,E,d,t) |
| Output : Labeled Multi-chain Graph |
| 0: Begin |
| 1: Remove all non-zero delay edges from the MDFG |
| 2: For each chain CH do |
| 3:     Compute the length L of CH |
| 4:     For each node starting from the last to the first do |
| 5:         Labeled the node by L |
| 6:         L=L-1 |
| 7:     End For |
| 8: End For |
| 9: End |

This technique starts by transforming the MDFG on Multi-chain Graph as described in Algorithm 4. Each chain represents a node succession where all interconnected edges between them are zero-delay. Each node is labeled by a level whose the value is greater than its predecessor node and smaller than its successor ones.

In the case of MDFG of IIR filter, the red integers above each node of Fig.8 represent the level values that are labeled after executing algorithm 4. Therefore, the multi-chain maximum length of $G$ is 4.

So, The technique proceeds to retime all the labeled nodes by a MD retiming function $(k - i) \times (r)$, as described in algorithm 5.

We present in Fig.8 the full parallelized graph of the IIR filter after applying the chained multidimensional retiming. The algorithm starts by finding the MDR function which is equal to $(1, -1)$. We note that all zero delay edges in the original MDFG are assigned by a delay vector equal to the MDR function.

Until now, no research works have been interested in comparing results provided by the techniques described above. The random choice of the scheduling vector does not allow defining the technique providing the optimal solution. However, based on their approaches, Chained MDR is generally more performing than the Incremental MDR. The first one consists in defining just one scheduling vector, which is executed in $O(|E|)$; while the second requires defining scheduling vector for each iteration of algorithm 4, which is executed in $O(|V|)$.

b) SPINE (Software PIpelining of NEsted loops) Multidimensional Retiming

This technique tries to provide a more optimal MDFG in terms of execution time and code size than those described above. It proceeds to remove all delays such as $(0, k)$ by merging them in a delay such as $(i, j)$. This modification is applicable only if the MDFG contains at least one edge with a delay equal to $(i, j)$ such as $i > 0$.

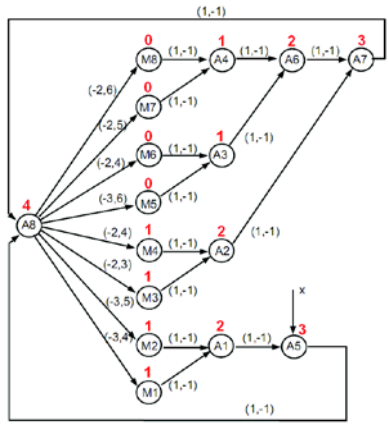

Fig.8. IIR Filter MDFG after Chained MDR.

It consists in finding a scheduling vector $s$ and a retiming function $r$ orthogonal to $s$, as described in algorithm 6, to provide a minimal overhead.

### 3.3 Multidimensional Retiming Constraints [5]

We describe in this paragraph the algorithmic constraints which must be taken into account to achieve full parallelism. These constraints come from the ratio between loop bounds and a number of MDR functions to apply.

For example, consider the multi-dimensional data flow graph in Fig.1; it is easy to verify that if $d(e4) = (0, k)$ where $k < 3$, retiming $D$ and $A$ by some vector $(0, p)$ will not satisfy the goal of re-distributing the delays among all edges in the graph. The same will happen if $d(e5) = (m, 0)$ where $m < 3$ for any retiming vectors of the form $(q, 0)$. Thus, if the loop has only one occurrence, i.e., the loop boundaries are both 1, then no parallelism can be obtained. This last constraint is equally applicable to a software or hardware implementation of the retimed loop.

This study begins by evaluating the constraints imposed by the limitation on the number of iterations comprising the loop. Since this limit is directly associated to the size of the iteration space, it is called spatial constraint and it is formally defined as follows.

ALGORITHM 6
SPINE-FULL ALGORITHM

Input : a realizable MDFG G =(V,E,d,t)
Output : a realizable MDFG $G_r$=(V,E,$d_r$,t) fully parallelized with minimum code size
1: Begin
2: If s=(1,0) is legal then
3:  Apply s=(1,0)
4: Else If s=(0,1) is legal and d(e)×(0,1)>=0 then
5:  Apply s=(0,1)
6: Else If s=(1,1) is legal and d(e)×(1,1)>=0 then
7:  Apply s=(1,1)
8: Else
9: Choose a legal scheduling vector s such as d(e)×s>=0, for any edge and $|s_x| + |s_y|$ is minimal
10: End

**Definition 1.** *Let a MDFG G contains k-level nested loop N, controlled by the set of indices $I = \{i_0, i_1, ..., i_k\}$, whose values vary, in unitary increments, in the range $L = \{l_0, l_1, ..., l_k\}$ to $U = \{u_0, u_1, ..., u_k\}$ where L is the set of lower boundaries for the indices and U is the set of maximum values, such as $l_j \le i_j \le u_j$, then the spatial constraint Sc of the loop is defined as:*

$$Sc = [(u_0 - l_0 + 1), (u_1 - l_1 + 1), ..., (u_k - l_k + 1)]$$

This definition allows establishing the relation between the maximum retiming operation and the spatial constraint according to the following lemma.

**Lemma 2.** *Given a k-level loop N with spatial constraint $Sc = [s_0, s_1, ..., s_k]$. The multi-dimensional retiming technique will be able to achieve full parallelism of the loop body instructions if the maximum retiming vector r applied to any node u, $r(u) = (r_0, r_1, ..., r_k)$ satisfies the following condition:*

$$r_j < s_j \text{ such as } 0 \le j \le k$$

### 3.4 Limitations of existing techniques

We have shown that nested loops can always be fully parallelized using MD retiming. The presented techniques of MD retiming are a polynomial time algorithm that fully parallelizes a given MDFG $G = (V, E, d, t)$ by selecting a legal schedule vector s with $s \times d(e) > 0$, $e \in E$, and a retiming vector $r$ where $r$ is orthogonal to $s$. Each MD retiming techniques presented above shows that the selected $s$ is a legal schedule vector for the retimed graph where $d_r(e) \ne d(e) \pm k.r$.

However, Multidimensional retiming techniques imply a large overhead of the generated code. It is caused by several aspects of loop transformation. First, the code size is increased because of the large code sections of the prologue and epilogue produced in all the loop dimensions. Second, the computation of the new loop bounds and loop indexes need to be recomputed [10].

Moreover, the execution of the prologue and epilogue section is not fully parallel, which requires a considerable period cycle number compared to that required by the loop body. Those disadvantages are aggravated in terms of the retiming vector value, and the number of the multidimensional retiming function.

Each Multidimensional Retiming techniques have a specific approach to choose the retiming function in order to decrease the overhead size. Chained and incremental Multidimensional retiming techniques proceed to chose a scheduling vector $s = (s.x, s.y)$, where $s.x + s.y$ is minimum. The SPINE technique tries to modify the MDFG with the intention of applying an MDR function that skews the minimum column-wise or/and row-wise. Despite providing an optimal solution, it is reliable only in the particular case of MDFG. In opposite cases, it applies the same approach than other techniques.

However, all existent techniques consist in retiming each node of the MDFG having an out-coming edge with zero-delay: if a path p as $d(p) = (0,0,...,0)$ is composed by n nodes, any technique applies $(n-1)$ MDR function to achieve full parallelism. But, overhead consequences are sudden after applying each MDR function: the more the number of MDR function increases, the more the consequences are dramatic.

As a result, the provided solution becomes very complicated and not sufficient to be implemented in embedded systems. Therefore, the existing MD retiming techniques, although achieving full parallelism, are not suitable for software nested loops.

## 4 THEORY OF OPTIMAL MULTIDIMENSIONAL RETIMING

In this section, we present the theoretical foundation of our proposal MDR technique "Optimal Multidimensional Retiming". It aims at minimizing MDR functions by exploring execution times and data dependency of nodes, while achieving full parallelism.

### 3.1 Principle

Multidimensional retiming techniques allow scheduling the MDFG with a minimal cycle period. For any path p $p: v_i \xrightarrow{e_m} v_{i+1} \xrightarrow{e_{m+1}} ... \xrightarrow{e_n} v_j$ of MDFG, they proceed to execute each node $v_k$ where $i \leq k \leq j$ in a period cycle separately. These approaches can be generalized into general-time cases [2] [4]. In fact, the computation nodes belonging to a data flow graph have not generally the same execution times. These depend on the kind of operation to be done; for example, a multiplication node needs usually more clock period than an adder node.

In this case, the minimal cycle period should be fixed differently. A cycle period represents a time interval leading to execute computation nodes. The minimal value of a period cycle can be defined as the smaller time interval that allows executing any node belonging to the MDFG. Thus, the minimal cycle period should be equal to the maximal execution time of node, as described in theorem 2 [1].

**Theorem 2.** Let $G = (V, E, d, t)$ a MDFG, the minimal value of cycle $c_{min}$ for the G graph is:

$$c_{min} = max\{t(v_i), v_i \in V\}$$

We choose to model the MDFG of IIR filter with different execution time of nodes such as of $t(M_i) = 3$ and $t(A_j) = 1$. The minimal cycle $c_{min}$ of this graph is equal to the execution time of the multiplication node.

Applying any MDR techniques to the IIR filter results in the fact that each iteration belonging to the original loop body is executed in 5 cycles. We illustrate in Fig.9 the static schedule of such an iteration after achieving full parallelism by the chained multidimensional retiming which is retimed by the function $r = (-1,1)$. The nodes belonging to the same iteration are modeled by gray circles.

Each gray node has not any data dependency with any other node executed in the same cycle period. However, these scheduling shows that provided data by some nodes are not consumed immediately. For example, nodes $A5$ and $A7$ are executed in just one time unit and their provided values are consumed two times units later. We conclude that cycle periods are not exploited optimally: (more than 66% of the cycle periods $(i-2, j+2)$, $(i-3, j+3)$ and $(i-4, j+4)$ are not used). However, they allow executing more than one node, due to the difference between execution time.

Let us try to execute nodes $A5$ and $A7$ in the same cycle period as nodes $A1, A2$ and $A6$, as schematized in Fig.10. The correspond MDFG with new delay values is illustrated in Fig.11. This transformation results in a legal MDFG that respects all conditions of lemma 1. Furthermore, it still keeps a fully parallelized execution, while preserving data dependency for the whole application. Compared to Fig.8, this transformation can be considered as decreasing the number of MDR functions by depriving nodes $A5$ and $A7$ to be retimed.

In fact, minimizing the number of MDR functions implies decreasing the overhead of the generated code. It consists in reducing the correspondent prologue, epilogue and the loop bound and index instructions. Moreover, it results in decreasing the number of cycle periods required to execute any iteration from 5 to 4, while respecting a fully parallelized execution. This minimization of cycle periods implies a similar minimization on the execution time of the whole application. Thus, this minimization of MDR functions leads to improve the performance of the provided full parallel solution.

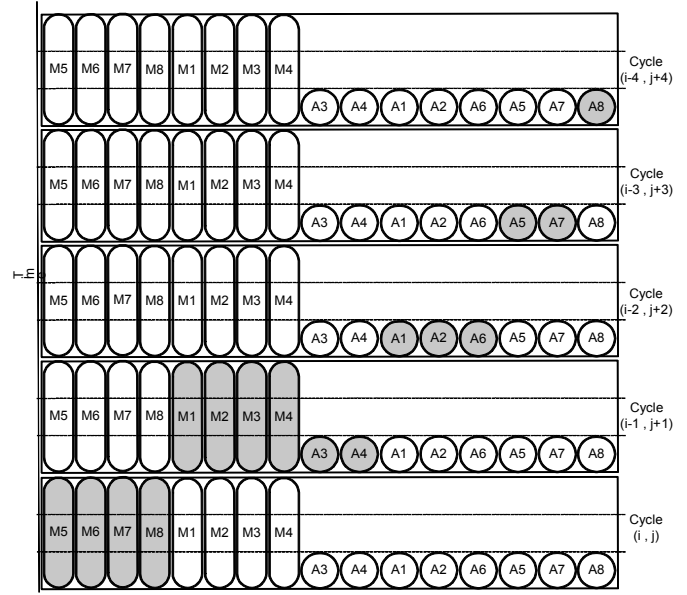

Fig.9. Iteration scheduling after chained MDR.

Such modification can be defined as applying the MDR to path p that can be composed of several nodes where $t(p)$ is smaller than $c_{min}$. The more the paths contain nodes, the more the MDR function number decreases. Thus, we propose a new MDR approach which consists in applying MDR function to a path of nodes, which can be executed in the same period cycle, instead of applying the MDR to each node separately. Our approach is based on selecting the path with the maximal nodes to achieve full parallelism with a minimal number of MDR functions. We start by computing the minimal period cycle $c_{min}$, extracting the paths with maximal nodes from the MDFG while keeping their execution in $c_{min}$, and applying the MDR function to the extracted paths.

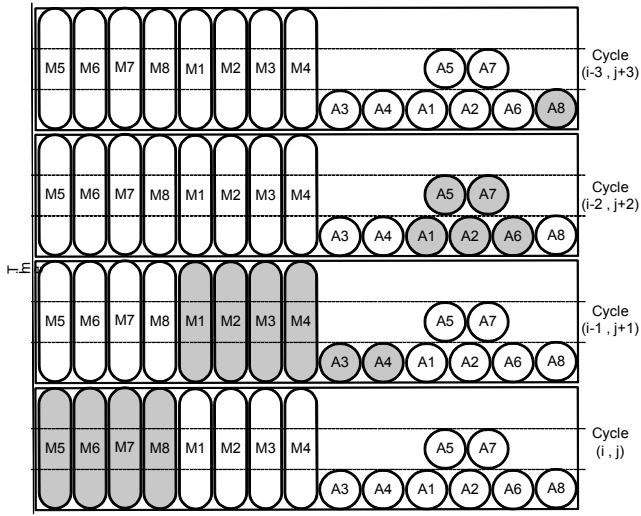

Fig.10. Iteration scheduling after collecting nodes in cycle (i-2, j+2).

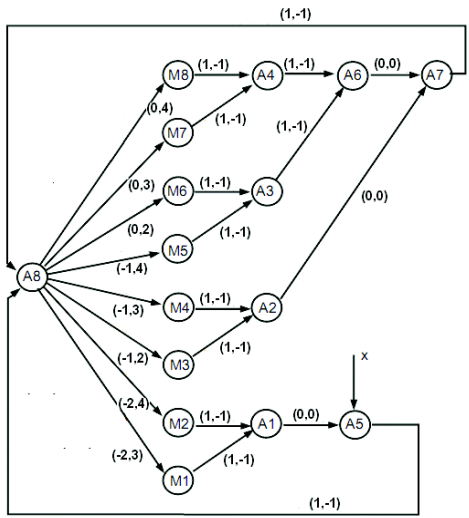

Fig.11. MDFG of Fig.10.

### 3.2 Basics concepts

Our technique consists in retiming a path of nodes that can be executed in the same cycle period. It means that those nodes are executed in the same iteration of the original loop body; i.e., all edges belonging to the path have zero-delay. We call such a kind of path "zero-delay path", which is indicated in theorem 3.

**Theorem 3.** Let $p: v_i \xrightarrow{e_m} v_{i+1} \xrightarrow{e_{m+1}} ... \xrightarrow{e_n} v_j$. p is zero-delay $d(p) = (0,...,0)$, if and only if :

$\forall e_l: v_k \rightarrow v_{k+1}, d(e_l) = (0,...,0);$ such as $e_l \in E, v_k \in V, k \in [i, (j-1)], l \in [m, n]$

Our technique proceeds to share the MDFG on zero-delay paths. It leads to maximize the node number in such a path, while being executed in the minimal cycle period. We define our idea of optimal multidimensional retiming as described in definition 2.

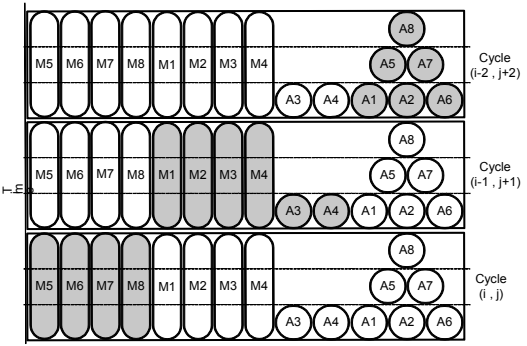

Fig.15. Iteration scheduling of MDFG shown in Fig.14.

**Definition 2.** Let $G = (V, E, d, t)$, $c_{min}$ the minimal period cycle of G. The optimal multidimensional retiming consists in retiming a set of path p follows :

1. $\max(card(p))$
2. $d(p) = (0,...,0)$
3. $t(p) \leq c_{min}$

Where $p: v_i \xrightarrow{e_m} v_{i+1} \xrightarrow{e_{m+1}} ... \xrightarrow{e_n} v_j$ as $v_k \in V \ \forall \ k \in [i, j]$ and card(p) is the number of nodes belonging to p such as $card(p) = j - i + 1$.

To preserve data dependency of MDFG, each p path should not contain any cycle in a way that no cycle between $v_p$ and $v_m$ where $\forall \ p, m \in [i, j]$ for each selected path $p: v_i \xrightarrow{e_m} v_{i+1} \xrightarrow{e_{m+1}} ... \xrightarrow{e_n} v_j$. Furthermore, multidimensional retiming consists in executing several paths in the same cycle period. Thus, such a path should not have any cycle between nodes that they belong to.

Our technique is based on retiming a zero-delay path. It means that all edges belonging to this path preserve the same delay (zero-delay). Only the delay of in-coming and out-coming edges of the whole path are changed. Thus, it means retiming the last node that it belongs to.

Referring to [2] and [3], the multidimensional retiming function is defined from edges having a non-zero delay of the MDFG. However, there is not constraint that requires applying the MDR function to nodes with non-zero in-coming edges. In the case of zero delay paths, it can be applied to any node belonging to them, as defined in theorem 4.

**Theorem 4.** Let $G = (V, E, d, t)$, and a zero delay path $p: v_i \xrightarrow{e_{(0,...,0)}} ... \xrightarrow{e_{(0,...,0)}} v_j$. If r is a legal MDR function of $v_i$, then r is a legal MDR function of $v_k$, where $v_k \in P$ and $k \in [(i+1), j]$.

**Proof.** a strict positive scheduling sub-space $S^+$ contains all scheduling vectors s where $d(e) \times s > 0$ for each $d(e) \neq (0,0,...,0)$, such as described in definition 2. This implies that $v_i$ and $v_k$ have the same sub-space $S^+$. But, a legal MDR r of $v_i$ is any orthogonal vector to s, where $s \in S^+$, as indicated in theorem 1. This means that a legal MDR r of $v_i$ is a legal MDR of $v_k$.

So, we provide an MDR function as indicated in theorem 2. This function is applied to the last node of the zero delay path to retime.

We conclude in section 3 that it is more sufficient to select an MDR function to apply it for all nodes to be retimed to achieve full parallelism, as described in corollary 1. Thus, we proceed by selecting every last node of a zero-delay path that will be retimed. Afterwards, we find a legal MDR function to be applied successively to the selected nodes while respecting data dependency of the MDFG.

ALGORITHM 8
OPTIMAL MULTIDIMENSIONAL RETIMING

```
Input  : a realizable MDFG G =(V,E,d,t)
Output : a realizable MDFG G_r=(V,E,d_r,t)
0:  Begin
1:  Find a legal MDR function r
2:  Provide the LMDFG and the maximal length M, as
    described in algorithm 7
3:  For i from M down to 1 do
4:      Select nodes v_k that are labeled by i
5:      For each v_k do
6:          Apply the MDR Retiming (i × r)
7:      End for
8:  End for
9:  End
```

### 3.3 Path Extraction

This step is based on exploring nodes belonging to the MDFG by testing their data dependency and execution time, to share them onto paths. We proceed by sweeping

ALGORITHM 7
LABELED MULTIDIMENSIONAL DATA FLOW GRAPH CONSTRUCTION

```
Input  : a realizable MDFG G =(V,E,d,t)
Output : Labeled Multidimensional Data Flow Graph
         (LMDFG), maximal length M
0:  Begin
1:  Compute c_min
2:  Extract all nonzero delay edges from G
3:  i =1
4:  Add the elements (v_j, t(v_j)) to R, such as v_j is a node
    without outcoming edge
5:  While R is not empty do
6:      For each v_j of R do
7:          Collect all predecessor of v_j
8:          For each predecessor v_p of v_j do
9:              If  t(v_p)+t(v_j) <= c_min and respect data dependency
                then
10:                 Add (v_p, t(v_p)+t(v_j)) to R
11:             Else
12:                 Add (v_p, t(v_p)) to N_E
13:             End If
14:         End for
15:     End for
16:     Label all nodes of N_E by i
17:     I= i+1
18:     R=N_E
19: End while
20: M = i
21: End
```

incrementally the MDFG from the opposite direction of edge. For each node, we verify that it respects the previous conditions, to execute it in the suitable cycle.

Our process consists on extracting the last node of each zero delay path that will be retimed, and labeled by an increasing order starting from 1. The result is illustrated in a "Labeled Multidimensional Data Flow Graph" (LMDFG) taken from the MDFG, as described in algorithm 7.

We proceed by exploring the MDFG in the opposite direction of data dependency. We start by extracting the non-zero delay edges from the MDFG and collecting all nodes $v_k$ without outcoming edge in $R$ list. For each node belonging to $R$, we define all the predecessor nodes $v_p$ to verify that can be executed in the same cycle time as nodes in $R$. For each $v_p$, if $t(v_p) + t(v_k) < c_{min}$ and if $v_p$ respects data dependency conditions described above, we consider $v_p$ as a node belonging to the path and we add it to the $R$ list. Else, $v_k$ is the last node of the previous path that should be retimed, and then we add then the node $v_k$ to $N_E$ list to label it. This test is repeated for all predecessor nodes of $R$ list element. The next step consists in labeling all the nodes of $N_E$ list by 1(first value of $i$), before replacing $R$ elements by $N_E$ elements. These steps are repeated until testing all nodes of MDFG.

We show in Fig.12 the LMDFG of IIR filter with $t(M_i) = 3$ and $t(A_j) = 1$. The first iteration of algorithm 7 consists in labeling the nodes $M1, M2, M3, M4, A3$ and $A4$ by 1. The last iteration assigns nodes $M5, M6, M7$ and $M8$ by 2. Therefore, the multi-chain maximum length is 2.

### 3.4 Optimal Multidimensional Algorithm

Our technique starts by finding a legal MDR r of MDFG, as indicated in theorem 2. After that, it provides the LMDFG and the maximal label by running algorithm 7. Then, it selects nodes with maximal label, and applies the MDR function $(i \times r)$. These steps are repeated by decreasing the label of the selected node until achieving retiming all the labeled nodes, as described in algorithm 8.

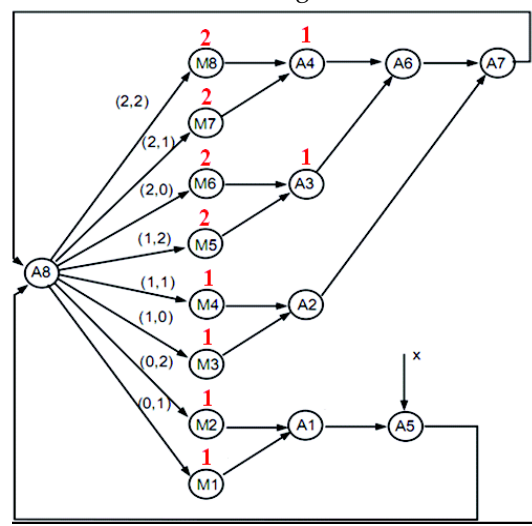

Fig.12. Labeled Multidimensional Data Flow Graph (LMDFG) of IIR filter.

As an example, we apply this algorithm to the MDFG of IIR filter. It starts by defining an MDR function $r = (1, -1)$, and providing the LMDFG shown in Fig.12. The first iteration of the algorithm retimes the nodes that are labeled by 2, by applying the retiming function $r = (2, -2)$ as illustrated in Fig.13. The second iteration retimes the other labeled nodes by $r = (1, -1)$, to provide the fully parallelized MDFG of IIR filter as shown in Fig.14. The nodes interconnected by a zero-delay edge are executed in the same cycle period. It means that nodes belonging to the same iteration in the original MDFG are executed in three cycle periods, as illustrated in the scheduling iteration of the fully parallelized MDFG in Fig.15.

## 5 EXPERIMENTAL RESULTS

In this section, we validate our MDR technique by comparing its provided MDFGs to those generated by the chained MDR. Four parameters are compared in our experimentation: the cycle period, the number of MDR functions, the execution time and the code size. We choose as an application the IIR filter graph and the wave digital filter after applying the Fettweis transformation [2], as illustrated in Fig.16. These two graphs are both composed of addition and multiplication nodes.

Our experiments consists in modeling each MDFG frequently in different cases of the execution time $t(v_i)$ where $v_i \epsilon V$, whose values are indicated in Table.1. We present in the last column the minimum cycle period whose values are defined as theorem 2.

For each MDFG in Table.1, we apply both chained MDR and optimal MDR techniques. Each one is characterized by a number of MDR functions applied to achieve full parallelism. The chained MDR generates the same fully parallelized MDFG with the same MDR function number, whatever the set of node execution time is.

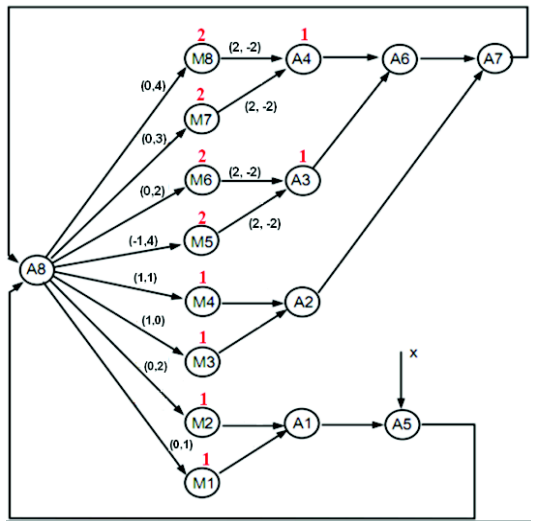

Fig.13. Retiming labeled nodes by 2.

Contrariwise, optimal MDR provides a specific MDFG for each MDFG of Table.1, with different MDR function numbers and retimed nodes. To guaranty a reliable comparison, we use the same MDR function for both techniques, which we apply the functions $r1 = (1, -1)$ and $r2 = (0,1)$ respectively to the IIR filter and wave digital filter. Table.2 illustrates the numbers of MDR functions

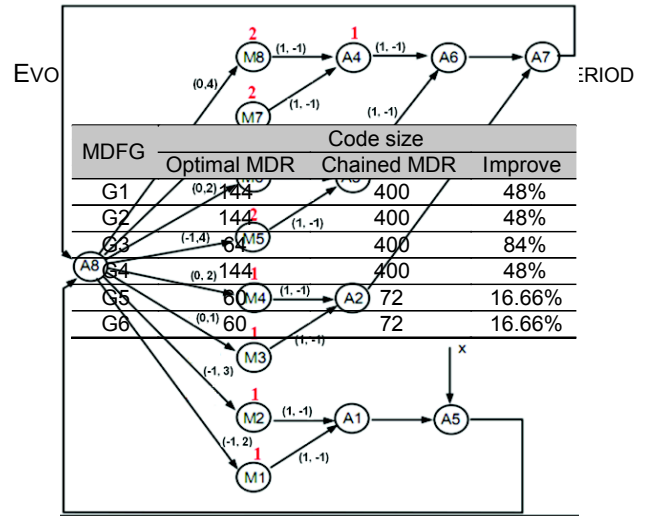

Fig.14. Full parallelized IIR filter after applying optimal MDR.

required to achieve full parallelism, for each MDFG, using both techniques.

After providing the fully parallelized MDFGs, we generate their respective algorithms, to extract their time and code size parameters. We present in Table.3 the values of the period cycle number and the execution time of each MDFG illustrated in Table.1 in terms of MDR techniques. The column "improve" presents the improvement of the execution time of result generated by our approach compared to those generated by the chained MDR, which accounts for an average improvement of 19.31%.

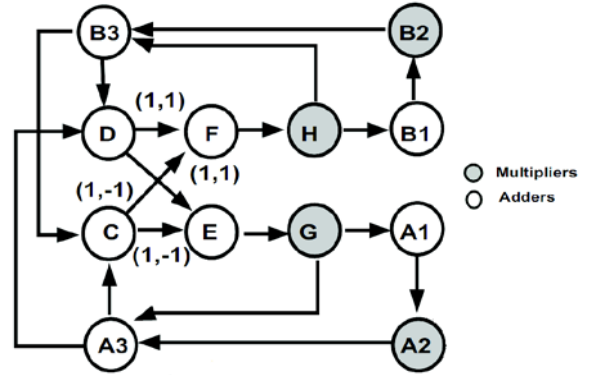

Fig.16. MDFG of wave digital filter

The code size of each MDFG provided in term of both of the two MDR techniques are shown in Table.4. Each value presents the instructions number of the loop body and the overhead caused by an MDR transformation. The code size values mention that our technique proposes an average improvement equal to 43.53% of the code size.

TABLE 1
CYCLE PERIOD IN TERMS OF NODE EXECUTION TIMES

| Application | MDFG | $t(A_i)$ | $t(M_j)$ | $C_{min}$ |
|---|---|---|---|---|
| IIR Filter | G1 | 1 | 2 | 2 |
| | G2 | 1 | 3 | 3 |
| | G3 | 1 | 4 | 4 |
| | G4 | 2 | 5 | 5 |
| WD filter | G5 | 1 | 2 | 2 |
| | G6 | 1 | 3 | 3 |

## 6 CONCLUSION

In this paper, we have proposed a new Multidimensional Retiming technique to achieve full parallelism of MDFGs. It allows providing an optimized MDFG compared to those provided by the existent techniques. It allows minimizing Multidimensional functions by exploring the execution times and data dependencies between the nodes.

In the section above, we have applied our technique and the chained MDR on different cases of MDFG. The results have shown that our technique generates a more efficient solution MDFG, than those generated by the chained MDR, in terms of cycle number, execution time and code size. We have concluded that our technique provides a more efficient solution, which allows respecting timing and code constraints while implementing the nested loop in embedded systems.

TABLE 2
EVOLUTION OF MDR FUNCTION NUMBER IN TERMS OF CYCLE PERIOD

| MDFG | MDR function number | |
|---|---|---|
| | Optimal MDR | Chained MDR |
| G1 | 2 | 4 |
| G2 | 2 | 4 |
| G3 | 1 | 4 |
| G4 | 2 | 4 |
| G5 | 1 | 2 |
| G6 | 1 | 2 |

As an optimization technique, we try in our future works to study using our MDR technique with other optimization approaches such as unrolling, loop fusion ... It consists in defining the applying order and the evolution of the performance parameter in terms of both approaches.

However, MDR techniques are based on scheduling the MDFG with a minimum cycle period. This period cycle value does not mean providing an MDFG with minimum execution time. We will be interested to extending our MDR technique to provide the adequate period cycle and a scheduling approach which allows providing MDFG with minimum execution time.

Also, in the case of real-time embedded system, the design consists in respecting the code size constraint, which should not be exceeded. This principle implies reducing execution time while achieving a limit value of the code size. Thus, based on the opposite evolution of the timing parameters and the code size in terms of MDR functions, we will be interested on proposing an optimization approach using the MDR technique: it requires finding the set of MDR functions that provide a retimed MDFG with a best ratio between the mentioned parameters.

TABLE 3
EVOLUTION OF CYCLE NUMBER AND EXECUTION TIME IN TERMS OF CYCLE PERIOD

| MDFG | Cycle number | | Execution time | | Improve |
|---|---|---|---|---|---|
| | Optimal MDR | Chained MDR | Optimal MDR | Chained MDR | |
| G1 | 258 | 316 | 516 | 632 | 18.35% |
| G2 | 258 | 316 | 774 | 948 | 18.35% |
| G3 | 229 | 316 | 916 | 1264 | 27.53% |
| G4 | 258 | 316 | 1290 | 1580 | 18.35% |
| G5 | 500 | 600 | 1000 | 1200 | 16.66% |
| G6 | 500 | 600 | 1500 | 1800 | 16.66% |

**Yaroub Elloumi** is a PHD student since September 2009, registered in both Paris-Est university (France) and Sfax university (Tunisia). He is a member of Institut Gaspard-Monge, unité mixte de recherche CNRS-UMLPE-ESIEE, UMR 8049. His research interests are High-level design of real-time system, optimization techniques, and High-level parameter estimation.

**Mohamed Akil** received his PhD degree from the Montpellier university (France) in 1981 and his doctorat d'état from the Pierre et Marie curie University (Paris, France) in 1985. He currently teaches and does research with the position of Professor at computer science department, ESIEE, Paris. He is a member of Institut Gaspard-Monge, unité mixte de recherche CNRS-UMLPE-ESIEE, UMR 8049. His research interests are Architecture for image processing, Image compression, Reconfigurable architecture and FPGA, High-level Design Methodology for multi-FPGA, mixed architecture (DSP/FPGA), System on Chip (SoC) and parallel programming of 2D/3D topological operators. Dr. Akil has more than 80 research papers in the above areas.

**Mohamed Hedi Bedoui** received his PhD degree from Lille University in 1993. He currently teaches with the position of Professor of biophysics in the Faculty of Medicine of Monastir (FMM), Tunisia. He is a member of Medical Technology and image processing team (TIM), UR 08-27. His research interests are real-time and embedded systems, image & signal processing and hardware/software design in medical field, electronic applications in biomedical instrumentation. He is the president of the Tunisian Association of Promotion of Applied Research.